\documentclass[slac_one]{revtex4}
\usepackage{graphicx}
\usepackage{fancyhdr}
\begin{document}

\title{Measurement of the pionium lifetime}

\author{M. Zhabitsky on behalf of the DIRAC Collaboration}
\email{zhabitsk@nusun.jinr.ru}
\affiliation{JINR, Dubna, RU141980, Russia}

\begin{abstract}
The goal of the DIRAC experiment
at CERN (PS212) is to measure
the pionium ($\pi^+\pi^-$~atom) lifetime.
The pionium lifetime is related to the $S$-wave
$\pi\pi$ scattering lengths difference $|a_0-a_2|$,
which was calculated within the framework of ChPT
with the 1.5\% precision.
Based on data collected in 2001-2003 on Ni targets
we report the significant improvement
in the measurement of the pionium
lifetime in the ground state.
Preliminary result of the pionium lifetime
has the precision of 11\%,
which corresponds to the measurement
of $|a_0-a_2|$ with the accuracy of 6\%.
\end{abstract}

\maketitle

\thispagestyle{fancy}

\section{INTRODUCTION}
Pionium ($A_{2\pi}$) is a Coulomb bound state
formed by a $\pi^+$ and a $\pi^-$~mesons.
Its lifetime of about $3\times 10^{-15}$~s
is dominated by annihilation
due to the strong interaction
($A_{2\pi}\rightarrow \pi^0 \pi^0$)~\cite{Uretsky61}.
Another annihilation channel $A_{2\pi}\rightarrow \gamma\gamma$
amounts on only $0.4$\%~\cite{Uretsky61,GasserPRep2008}.
Pionium annihilates predominantly from $nS$-states.
In pionium the two pions interact
at very low relative momentum~$Q$
(of the order of $0.5\:\text{MeV}/c$)
and in a well defined quantum states,
therefore its lifetime in $nS$-states is directly related
to the $S$-wave $\pi\pi$~scattering lengths $a_0$ and $a_2$,
with isospin 0 and 2 respectively~\cite{Uretsky61,GasserPRD2001}:
\begin{equation}
 \tau_{nS}^{-1} =
  \Gamma_{A_{2\pi}(nS)\to\gamma\gamma} +
  \Gamma_{A_{2\pi}(nS)\to\pi^0\pi^0},\qquad
  \Gamma_{A_{2\pi}(nS)\to\pi^0\pi^0} = 
\frac{2\alpha^3 p^*}{9 n^3}\,
|a_0-a_2|^2m_{\pi^+}^2 (1+\delta),
\label{eq:tau_a0a2}
\end{equation}
where $\alpha$ is the fine-structure constant,
$p^*$ is the $\pi^0$ momentum in the atomic rest frame.
The term~$\delta$ accounts for QED and QCD corrections
and is a known quantity:
$\delta=(5.8\pm 1.2)\times 10^{-2}$~\cite{GasserPRD2001}.
The difference $(a_0-a_2)$
of $S$-wave $\pi\pi$~scattering lengths
was calculated within the framework
of Standard Chiral Perturbation Theory
with the $1.5$\% precision~\cite{ColangeloNPB2001}:
$a_0-a_2=(0.265\pm 0.004)\times m_\pi^{-1}$.
Generalized Chiral Perturbation Theory
though allows for larger $a$-values~\cite{Knecht}.
Thus the measurement of pionium lifetime by DIRAC
provides the test of concepts
of low-energy QCD
through a constraint on properties of pion-pion interaction.

\begin{figure}
\centering
\includegraphics*[width=0.7\textwidth]{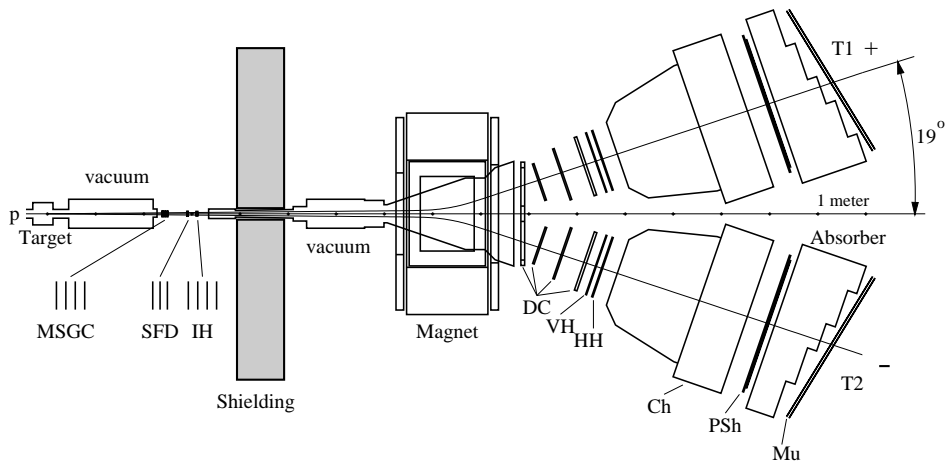}
\caption{Schematic top view of the DIRAC spectrometer.
Vertex detectors between target and the magnet:
4~planes of microstrip gas chambers (MSGC),
3~planes of scintillating fiber detectors (SFD) and
4~planes of scintillation ionization hodoscope (IH).
Downstream the dipole magnet:
drift chambers (DC),
vertical (VH) and horizontal (HH) scintillation hodoscopes,
nitrogen Cherenkov counters (Ch),
preshower (PSh) and
behind the iron absorber muon hodoscope (Mu).}
\label{fig:setup}
\end{figure}

In DIRAC pioniums are originated
from collisions of $20$ or $24~\text{GeV}/c$ protons
of the CERN PS
with a typically 100~$\mu$m thick Ni foil.
The experiment uses a double-arm magnetic
spectrometer~\cite{DIRACNIM} (Fig.~\ref{fig:setup}),
which is optimized to detect $\pi^+\pi^-$~pairs
with a small relative momentum~$Q$
in their center-of-mass system.
The resolution on the components
of pairs relative momentum~$Q$ is $\lesssim 0.5\:\text{MeV}/c$,
which is about the pionium Bohr momentum.

The method of the pionium lifetime measurement~\cite{Nemenov85}
used in the DIRAC experiment
is as follows.
Pioniums are produced in $nS$-states in proton-Ni collisions.
Probability of pionium production
depends on the principal quantum number as $n^{-3}$.
While crossing the target pionium can be excited or ionized
(form an unbound pair of a $\pi^+$ and a $\pi^-$~mesons)
due to Coulomb interaction with target atoms.
Pairs of a $\pi^+$ and a $\pi^-$~mesons from atom break-up
(``\textit{atomic pairs}``)
can be identified through the analysis of $Q$-spectra
thanks to their distinct low-$Q$ distribution.
Ionization of pionium is a process competitive
to its annihilation.
Therefore the probability of pionium ionization in the target
is a unique function of its lifetime,
which is calculated
with sufficient precision~\cite{Cibran2003,Zhabitsky2008}.

In proton-Ni collisions, besides pioniums,
may emerge low-$Q$ $\pi^+\pi^-$~pairs (``\textit{Coulomb-correlated pairs}''),
which also undergo Coulomb interaction in the final state
but don't form a bound state.
The number~$N_A$ of \textit{produced} atoms is therefore
linked to the number~$N_{CC}$ of Coulomb-correlated pairs
with small~$Q$,
e.g. at $Q\leq 2\:\text{MeV}/c$
$N_A/N_{CC}=k_{\text{theor}}=0.615$.
Coulomb-correlated $\pi^+\pi^-$~pairs exhibit an enhancement at low~$Q$
with respect to the phase space~\cite{Sakharov48}.
Through the analysis of low-$Q$ spectra
collected by the spectrometer
we are able to reconstruct
the number~$n_A$ of \textit{ionized} atoms
and the number~$N_{CC}$ of {CC-pairs},
therefore the probability of pionium ionization in the target is measured:
\begin{equation}
 P = \frac{n_A}{\epsilon\, k_{\text{theor}} N_{CC}},
\end{equation}
where $\epsilon$ takes into account relative efficiencies
to detect \textit{atomic} and \textit{Coulomb-correlated} pairs
obtained by event simulation using Monte Carlo method.

Finally in inclusive $\pi^+\pi^-$~pairs
there is a background of \textit{non-Coulomb} (NC) pairs, 
if either or both of pions emerge
from long-lived sources
and thus NC-pions do not interact in the final state.

\section{THE MEASUREMENT}

The details of event reconstruction,
data selection and analysis
have been published in~\cite{DIRAC_pipidetection,DIRACPLB2005},
where the first measurement of the pionium lifetime
$\tau_{1S}=\left(2.91_{-0.62}^{+0.49}\right)\times 10^{-15}$~s
based on statistics of about 6500~ionized atoms collected by DIRAC
in 2001 on Ni target was reported.

\begin{figure}
\begin{minipage}{0.48\textwidth}
\centering
\includegraphics*[scale=0.8]{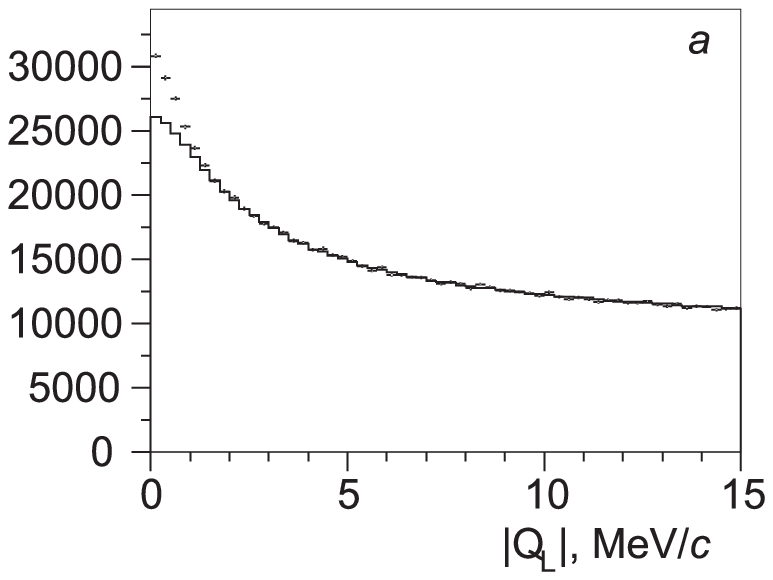}
\end{minipage}\hfill%
\begin{minipage}{0.48\textwidth}
\centering
\includegraphics*[scale=0.8]{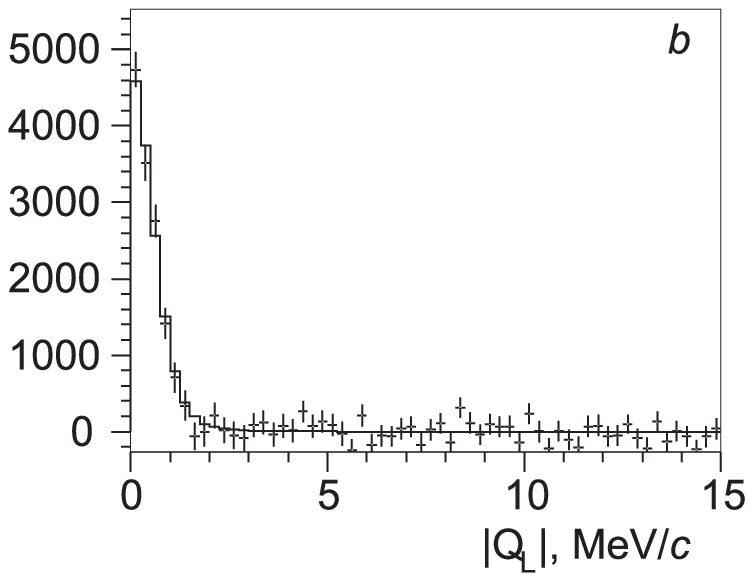}
\end{minipage}\hfill%
\caption{\textit{a)}~--- Measured distribution of inclusive $\pi^+\pi^-$~pairs:
experimental points are shown with
the sum of CC and NC distributions (solid line).
\textit{b)}~--- measured distribution after subtraction
of the sum of CC and NC distributions,
simulated distribution of atomic pairs is shown by the solid line.}
\label{fig:inclusiveQL}
\end{figure}

In 2001--2003 DIRAC has accumulated more than $15\times 10^3$
atomic pairs.
The data was collected on two Ni foils,
one was $94\:\mu$m thick and another of $98\:\mu$m thickness.
Momenta of proton beam was either $24$
(2001 and a part of 2002) or $20\:\text{GeV}/c$
(another part of 2002 and 2003).
Here we report preliminary results of analysis,
based on a part of the accumulated statistics:
signals from only two of vertex detectors (SFD and IH)
were taken into account and
only events with low background conditions were used.
Fig.~\ref{fig:inclusiveQL},\,\textit{a} shows the measured distribution
of inclusive $\pi^+\pi^-$~pairs on $|Q_L|$,
which is the magnitude of $Q$-component along
the pairs momentum.
While the distribution of NC-pairs on $Q_L$ is uniform,
the Coulomb enhancement
below $10\:\text{MeV}/c$ reveals the number $N_{CC}^{\text{exp}}$
of Coulomb-correlated pairs in the sample.
After subtraction of CC and NC pairs
one finds the signal of atomic pairs
shown in fig.~\ref{fig:inclusiveQL},\,\textit{b}.

Results of the preliminary analysis of experimental data collected
in 2001--2003 on Ni targets are summarized in Tab.~\ref{tab:pbr_exp}.
The results were obtained from ($Q_L$, $Q_T$) 2D-fit.
The data sample was divided into three periods
with different target thicknesses and different proton beam momenta,
thus these periods are characterized
by distinct dependencies $P_i(\tau)$
of the probability of pionium ionization in the target
as a function of its lifetime.

\begin{table}
\centering
\caption{Number of reconstructed atomic pairs ($n_A^{\text{exp}}$),
CC-pairs ($N_{CC}^{\text{exp}}$) and the probability
of ionization~$P_i$ for different sets of experimental data.
The cuts on $Q$-components are
$|Q_{x,y}|<4$ and $Q_L<15\:\text{MeV}/c$}
\label{tab:pbr_exp}
\begin{tabular}{|c|c|r|r|r|}\hline
\rule{0pt}{12pt}%
Proton &
Target &
\multicolumn{1}{c|}{$n_A^{\text{exp}}$} &
\multicolumn{1}{c|}{$N_{CC}^{\text{exp}},\ 10^3$} &
\multicolumn{1}{c|}{$P_i$} \\
momenta, $\text{GeV}/c$ &
thickness, $\mu$m&
&&\\ \hline
\rule{0pt}{12pt}%
24 & 94 & $4438\pm 192$ & $328.4\pm 3.2$ & $0.459\pm 0.022$ \\
24 & 98 & $5011\pm 274$ & $402.1\pm 4.0$ & $0.421\pm 0.025$ \\
20 & 98 & $3851\pm 192$ & $291.7\pm 3.1$ & $0.446\pm 0.025$ \\ \hline
\multicolumn{2}{|c|}{\rule{0pt}{12pt}%
Total}& $13300\pm 386$ &$1022.1\pm 6.0$ & \\ \hline
\end{tabular}
\end{table}

\begin{table}
\centering
\caption{Summary of systematic uncertainties
in the measurement of the probability of pionium ionization in the target}
\label{tab:syst}
\begin{tabular}{|l|c|}\hline
\rule{0pt}{12pt}\mbox{}  & $\sigma(P_i)$ \\ \hline
\rule{0pt}{12pt}%
Multiple scattering                         & 0.006\\
Admixture of $K^+K^-$ and $\bar{p}p$        & 0.0005\\
Finite size effects                         & ${}_{-0.006}^{+0.000}$\\
Correction on momenta resolution            & 0.008\\
Double track resolution in vertex detectors & 0.003\\
Background hits in vertex detectors         & 0.0012\\
Trigger                                     & 0.0013\\ \hline
\rule{0pt}{12pt}%
Total
                   & ${}_{-0.012}^{+0.010}$\\ \hline
\end{tabular}
\end{table}

The systematic errors are summarized in Tab.~\ref{tab:syst}.
Thanks to the dedicated measurement of scattering angle
in any of scatterers
from the target foil towards Drift Chambers~\cite{note0806}
the error due to multiple scattering was reduced
by a factor of 2 in comparison to the first publication~\cite{DIRACPLB2005}.
Through time-of-flight measurement by vertex detectors
and vertical hodoscopes downstream the magnet
the amount
of $\bar{p}p$~\cite{pp_admixture} and $K^+K^-$-pairs~\cite{KK_admixture}
were estimated,
thus this uncertainty has been almost removed
with respect to the previous publication.
For the data analysis we assumed point-like production
of Coulomb-correlated $\pi^+\pi^-$~pairs,
but small part of CC-pairs originates
from sufficiently long-lived resonances,
therefore the uncertainty due to a finite size
of their production region was estimated~\cite{lednicky}.
In order to reproduce in the Monte Carlo simulation the width
of detected $\Lambda\to p\pi^-$ decays,
the correction on momenta resolution was introduced~\cite{note0717}.
Smaller systematic errors
are due to uncertainties in track reconstruction in vertex detectors
and trigger inefficiency.
More details can be found in an in-depth study~\cite{note0804}.

The measurements of pionium ionization probability from Tab.~\ref{tab:pbr_exp}
are correlated due to the systematic uncertainty
which is common for all measurements.
Let $G$ be the error matrix:
$G_{ij}=\sigma_{\text{sys}}^2 + \delta_{ij}{\sigma_{\text{stat}}}_i^2$,
where $\sigma_{\text{sys}}$ is the systematic uncertainty
and ${\sigma_{\text{stat}}}_i$ is the statistical uncertainty
in the measurement $P_i$ of the probability of ionization.
With relations $P_i(\tau_{1S})$,
calculated for experimental conditions of different periods,
the estimation of the lifetime
together with its uncertainties
is achieved~\cite{note0807} by the maximum likelihood method,
where the likelihood function reads
\begin{equation}
L(\tau_{1S}) = \exp\left[-\frac{1}{2}
 \sum_{ij}\left(P_i-P_i(\tau_{1S})\right)
 \left(G^{-1}\right)_{ij}
 \left(P_j-P_j(\tau_{1S})\right)
\right].
\end{equation}

\begin{table}[h]
\caption{Comparison of the precision of the results based
on the preliminary analysis of 2001--2003 statistics
to the precision of the published result~\cite{DIRACPLB2005}}
\label{tab:results}
\begin{tabular}{|l|c|c|}\hline
\rule{0pt}{12pt}%
Period of data-taking         & \multicolumn{1}{|c|}{2001}
                                & \multicolumn{1}{c|}{2001-2003}\\ \hline
$n_A$                         & $6530 \pm 294$   & $13300\pm 386$\\ \hline
\rule{0pt}{12pt}%
statistical uncertainty in $\tau_{1S}$, $10^{-15}\:$s &
  ${}^{+0.45}_{-0.38}$ & ${}_{-0.23}^{+0.25}$\\[2pt]
systematic uncertainty in $\tau_{1S}$, $10^{-15}\:$s &
  ${}^{+0.19}_{-0.49}$ & ${}_{-0.19}^{+0.19}$\\[2pt]
total uncertainty in $\tau_{1S}$, $10^{-15}\:$s &
  ${}^{+0.49}_{-0.62}$ & ${}_{-0.30}^{+0.31}$\\ \hline
\rule{0pt}{12pt}%
total uncertainty in $|a_0-a_2|$, $m_{\pi^+}^{-1}$ &
  ${}_{-0.020}^{+0.033}$ & ${}_{-0.014}^{+0.016}$\\[2pt] \hline
\end{tabular}
\end{table}

In Table~\ref{tab:results} we compare the precision
of the preliminary results based on 2001--2003 statistics
to the precision of the published result~\cite{DIRACPLB2005}.
The statistical errors are not symmetrical
because of a non-linearity of $P(\tau_{1S})$ relation.
The analyzed statistics is twice as big
as the sample, used in the previous publication.
There is a better understanding of major sources of systematic uncertainties.
Thanks to the dedicated measurements the systematic error
has been reduced.
The total uncertainty in the measurement of the pionium lifetime
is about 11\%,
which corresponds to the measurement
of $|a_0-a_2|$ with the accuracy of 6\%.
Further improvements are expected,
when we will use all statistics accumulated
in 2001--2003 on nickel targets
and all vertex detectors will be incorporated
into the track reconstruction.

\end{document}